\newcommand\hii{\ion{H}{2}}

\documentstyle[11pt,aaspp4]{article}
\lefthead{GARNETT \& KOBULNICKY}
\righthead{AGE-METALLICITY RELATION}
\received{11 August 1999}
\revised{29 October 1999}
\accepted{17 November 1999}
\slugcomment{To appear in the April 10, 2000 Astrophysical Journal}
\begin{document}
%\title{RE-EXAMINING THE SOLAR NEIGHBORHOOD AGE-METALLICITY RELATION}
\title{DISTANCE DEPENDENCE IN THE SOLAR NEIGHBORHOOD AGE-METALLICITY RELATION}
\author{D. R. Garnett}%\altaffilmark{1} 
\affil{Steward Observatory, University of Arizona, 933 N. Cherry Ave., 
Tucson AZ 85721\\ e-mail: dgarnett@as.arizona.edu}
\and 
\author{H. A. Kobulnicky\altaffilmark{1,2} }
\affil{Lick Observatory, University of California at Santa Cruz, 
Santa Cruz, CA 95064}
\altaffiltext{1}{Current address: Astronomy Department, University of Wisconsin, 
Madison, WI 53706; e-mail: chip@astro.wisc.edu}
\altaffiltext{2}{Hubble Fellow} 
\begin{abstract}

The age-metallicity relation for F and G dwarf stars in the solar 
neighborhood, based on the stellar metallicity data of Edvardsson et 
al. (1993), shows an apparent scatter that is larger than expected 
considering the uncertainties in metallicities and ages. A number of 
theoretical models have been put forward to explain the large scatter. 
However, we present evidence, based on Edvardsson et al. (1993) data,
along with Hipparcos parallaxes and new age estimates, that the scatter 
in the age-metallicity relation depends on the distance to the stars in 
the sample, such that stars within 30 pc of the Sun show significantly 
less scatter in [Fe/H]. Stars of intermediate age from the Edvardsson 
et al. sample at distances 30-80 pc from the Sun are systematically
more metal-poor than those more nearby. We also find that the slope of 
the apparent age-metallicity relation is different for stars within 30 
pc than for those stars more distant. These results are most likely an 
artifact of selection biases in the Edvardsson et al. star sample. We 
conclude that the intrinsic dispersion in metallicity at fixed age is 
$<$ 0.15 dex for field stars in the solar neighborhood, consistent with 
the $<$ 0.1 dex for Galactic open star clusters and the interstellar medium. 

\end{abstract}

\section{Introduction}

The age-metallicity relation (AMR) for stars, coupled with the stellar
metallicity distribution and the star formation history, is a 
fundamental constraint on models for the chemical evolution of the 
solar neighborhood, providing the time history of the enrichment of 
the interstellar medium. Defining this relation, however, has not 
been a trivial task, as it requires obtaining the ages, distances, 
metallicities, and kinematics for a large sample of stars. The 
age-metallicity relation established by Twarog (1980) \markcite{tw80}
was a key constraint on chemical evolution models for the solar 
neighborhood. However, this study lacked kinematic information for
the stars, and thus on the amount of contamination by stars not 
born in the solar neighborhood (such as thick disk stars, whose
connection to galactic evolution is uncertain at present).

More recently, Edvardsson et al. (1993; hereafter Edv93)\markcite{edv93} 
published abundances for numerous heavy elements in field F and G dwarf 
stars having kinematic information. These data have provided a wealth of 
information on abundances and abundance ratios as a function of time and
kinematics in the galactic disk. The sample of stars chosen had a variety 
of photometric information and space velocities, allowing them to be placed 
on the HR diagram and in the appropriate kinematic population. One surprising 
result from this study was that the AMR derived by Edv93 showed much 
greater scatter (inferred to be 0.6-1.0 dex by some papers) than could be 
attributed to observational uncertainties. This result suggested that 
chemical evolution in the solar neighborhood has been highly inhomogeneous 
over time.

A number of theoretical explanations for the scatter in the local AMR have 
been proposed, including: radial diffusion of 
stellar orbits (Fran\c cois \& Matteucci 1993\markcite{fm93}; Wielen, 
Fuchs, \& Dettbarn 1996\markcite{wfd96}); episodic infall of metal-poor 
gas (Edv93; Pilyugin \& Edmunds 1996\markcite{pe96}); and sequential or 
stochastic enrichment by stellar populations (van den Hoek \& de Jong 
1997\markcite{vhdj97}; Copi 1997)\markcite{copi97}. Which mechanism 
might be most important is unknown, but there is no lack of explanations. 
On the other hand, the inference of large scatter is inconsistent with 
abundance measurements in nearby spiral and irregular galaxies (e.g., 
Kennicutt \& Garnett 1996\markcite{kg96} and Kobulnicky \& Skillman 
1996\markcite{ks96}), and in the local ISM (Meyer, Jura, \& Cardelli 
1998\markcite{mjc98}), which show that dispersions in ISM abundances 
are rather small on kiloparsec scales or less. It is difficult to 
understand how a largely homogeneous ISM could give rise to a large 
dispersion in stellar metallicities. The apparent dispersion in the
Edv93 data is inconsistent with the smaller dispersion derived by Twarog
(1980) as well.

Therefore, it seems appropriate to re-examine the AMR. Edv93 warned that 
their star sample was not an unbiased sample (see also Nissen 
1995) and thus should be used cautiously in interpreting the AMR. We 
will demonstrate below that there is a systematic dependence 
in the amount of scatter in the AMR on the properties of the stars, 
in particular on stellar distance. We will 
conclude that the intrinsic scatter in the AMR is 
smaller than has sometimes been inferred.

\section{The Stellar Sample}

The Edv93 metallicity sample consisted of 189 somewhat evolved 
F and G stars within 80 pc of the sun. The sample was selected to 
have roughly equal numbers of stars in nine metallicity bins from [M/H] 
= +0.2 to $-$0.9, where [M/H] is the logarithmic metallicity relative 
to solar. In order to obtain such sampling, Edv93 had 
to observe fainter and more distant stars to obtain a sufficient number 
of stars in the low-metallicity bins.

Of this sample, eleven stars are spectroscopic binaries, which can have
larger uncertainties in distances and ages; another eleven stars had
large uncertainties in their proper motions. We have excluded these from 
our analysis. Seven other stars had no age estimates and were also excluded. 
We have retained the stars labeled `hook' stars by Edv93, although 
their ages may be systematically underestimated by up to 0.15 dex.

Not all of the stars in the Edv93 sample had trigonometric parallaxes in 
1993, and so they relied on distances based on their Str\" omgren photometry. 
Since then, parallaxes from the Hipparcos catalog (ESA 1997) have become 
available for all of these stars. Ng \& Bertelli (1998)\markcite{nb98} have 
published revised ages for the stars based on the Hipparcos parallaxes and 
the more recently computed stellar evolution tracks of Bertelli et al. 
(1994)\markcite{bbcfn94}. Comparison of the revised stellar ages with 
the Edv93 ages showed little systematic difference (Ng \& Bertelli 
1998\markcite{nb98}), indicating that age uncertainties are not the 
main source of the scatter in age vs. metallicity. However, there 
is a slight reduction in the scatter when the Hipparcos distances and 
Ng \& Bertelli (1998)\markcite{nb98} ages are used; we will therefore 
base the following discussion on the revised ages from Tables 5 
and 6 of Ng \& Bertelli (1998)\markcite{nb98} and Hipparcos distances.

\section{Distance Dependence of the Scatter in Metallicity}

Figure 1 shows the AMR plot for [Fe/H] based on our selected subset 
of the Edv93 stars. Assuming the quoted observational uncertainties of 
$\pm$0.1 dex in [Fe/H] and age, a Monte Carlo analysis indicates that an 
additional Gaussian dispersion in [Fe/H] of $\pm$0.15-0.2 dex beyond the 
observational scatter is required to account for the observed scatter. 

We were originally concerned that uncertainties in the distances to the
stars could introduce errors into their ages, and thus could cause
artificially large scatter in the AMR. Therefore, we 
examined the scatter as a function of distance to the stars. We made a 
simple division of the sample into two groups: a nearby group with 
distances less than 30 parsecs from the sun (89 stars), and a more 
distant set containing the remaining 71 stars with distances greater 
than 30 parsecs (which extends out to 80 parsecs). Figure 2 shows 
the age-metallicity diagrams for the two groups of stars. The comparison 
is remarkable: the nearest stars show an obvious reduction in scatter 
in [Fe/H] at a given age than the full sample of Figure 1, especially for 
the intermediate ages (log $\tau$ between 0.5 and 0.9 in Gyrs). In fact,
Figure 2 reveals a striking asymmetry in the metallicity distribution for
the nearby and more distant stars. While for the nearby stars [Fe/H] rises
steeply with decreasing age and then levels off for the younger stars, for
the more distant stars a more gradual increase in [Fe/H] appears to be the 
case. To show that this difference is not a simple statistical fluctuation, 
we examine the stars in the range 0.5 $<$ log $\tau$ $<$ 0.9, where the more 
distant sample shows a dearth of metal-rich stars and an enhancement in 
the number of metal-poor stars compared to the nearby sample. We show the 
distributions of [Fe/H] for the near and far stars within this age range 
in Figure 3. A Mann-Whitney test rejects the hypothesis that these two 
samples come from populations with the same mean [Fe/H] at the 99.99\% 
confidence level, while a Kolmogorov-Smirnov test indicates that the 
probability that these two samples are drawn from the same population is 
only 8.3$\times$10$^{-4}$. Thus, it appears that the difference between
these two groups of stars is highly significant. (We see the same patterns 
in plots for other elements as well.) The 
difference suggests that the more distant sample may include stars whose 
chemical properties do not reflect the evolution of the disk in the solar 
neighborhood. This is plausible since the more distant stars sample a volume 
that is 18 times larger than the stars within 30 pc.

We explore the properties of the two star samples further. Figure 4 plots 
[Fe/H] vs. the mean stellar orbital radius $R_{mean}$ (from Edv93), Figure 
5 shows [Fe/H] vs. the maximum height $Z_{max}$ of each star above 
the Galactic plane (one star, HD148816, lies outside Fig. 5 with [Fe/H] 
= $-$0.74 and $Z_{max}$ = 5.44 kpc.), and Figure 6 plots [Fe/H] vs. 
orbital eccentricity. (We have not attempted to re-derive orbital 
parameters for the stars based on the new parallax results. Although 
there may be significant changes for a few stars, for the vast majority
of these stars distances have changed only slightly and so the orbits will 
also change negligibly.) Two things can be discerned from these plots. First, 
there is a tight cloud of disk stars with mean orbital radii between 7.0 
and 8.5 kpc, orbit eccentricities $<$ 0.15, and [Fe/H] $>$ $-$0.5. Second, 
most of the metal-poor stars in the sample have eccentric orbits that range 
far from the solar radius and away from the galactic plane. Edv93 also noted 
the increase in the vertical velocity dispersion for old, metal-poor stars 
in their sample (see their Fig. 16). It can be inferred from this that the 
sample is contaminated by thick disk stars and other stars from outside the 
solar circle. 

The differences between the upper and lower panels of Figure 2 largely reflect 
the sample selection criteria used by Edv93. Metal-poor stars are more rare 
than metal-rich ones in the solar neighborhood. Thus, in order to have roughly 
equal stars in each metallicity bin, the metal-poor stars will be fainter 
and more distant, on average. Second, Edv93 selected stars which were at 
least 0.4 magnitudes above the main sequence but within the temperature range 
5600-7000 K. Finding young stars that meet these criteria is more difficult
than finding older stars. Therefore, the young stars will also tend to be
more distant. The third difference is the fact that the stars in the outer
shell with 0.5 $<$ log $\tau$ $<$ 0.9 are systematically more metal-poor 
than stars of the same ages in the inner shell. We suspect that this is
also likely a result of the selection criteria, but this is more difficult
to understand without modeling the selection biases. A comparison of the 
kinematics of the two sets could prove interesting, but is beyond the scope 
of this paper. Figure 2 makes it clear that 
this is not a randomly selected sample of stars. If the inner and outer 
circles were fair samples of the solar neighborhood metallicity distribution, 
Figs. 2(a) and 2(b) should show similar age-metallicity relationships.

Finally, we compare the [Fe/H] distribution for the stars in the Edv93 
sample with the volume-limited sample of G and K dwarfs from Favata, 
Micela, \& Sciortino (1997)\markcite{fms97}. Favata et al. derived [Fe/H] 
for a random selection of 92 stars from the Gliese catalog of nearby stars. 
There are eight stars in common between Edv93 Favata et al.; the mean 
difference in [Fe/H] is 0.03$\pm$0.03 dex, suggesting no significant
systematic difference in the two metallicity scales. However, Favata et al.
noted a peculiarity in their [Fe/H] distribution, in that the cooler K dwarfs
showed a higher mean and smaller dispersion in [Fe/H] than the G dwarfs. For
fair comparison, therefore, we restrict our discussion to the 39 Favata et al.
stars with $T_{eff}$ $>$ 5600 K, the lower $T_{eff}$ limit of the Edv93 sample,
to be compared to the Edv93 stars within 30 pc of the Sun. 

We plot the [Fe/H] distributions of these two samples as histograms in Figure 
7. The Edv93 stars are clearly skewed toward lower metallicities compared 
to the Favata et al. sample, with the Edv93 sample missing the most 
metal-rich stars found by Favata et al., while showing an excess of stars 
in the range $-$0.1 $>$ [Fe/H] $>$ $-$0.4. Again, this could be a result of 
the metallicity selection in the Edv93 sample. It is probably not possible to 
say at present if the volume-limited sample has a smaller dispersion than 
the Edv93 sample, given the small numbers of stars involved. Unfortunately, 
the Favata et al. stars lack kinematic data, so a more detailed comparison 
is not possible at present.

\section{Discussion}

It is apparent from this exercise that determining the shape and dispersion
in the age-metallicity relation is not particularly straightforward, even 
for a high-quality data set such as the Edv93 sample. One must be careful 
to account for kinematically distinct populations, sample selection, and 
abundance peculiarities 
to derive a representative solar neighborhood sample. Although diffusion of
stars from outside the solar circle does contribute somewhat to the scatter
in abundances, it does not account for most of it, as was suggested by Wielen,
Fuchs, \& Dettbarn (1996). That this is the case can be inferred by comparing
the Edv93 AMR with that from Twarog (1980). The measured dispersion in [Fe/H] 
at every age bin in the Twarog sample is smaller than that measured in the
Edv93 sample; the dispersion in [Fe/H] in the Twarog sample ranges from 0.06 
to at most 0.18 dex, compared with the 0.24 dex determined by Edv93. If stellar 
diffusion is responsible for most of the scatter in the Edv93 AMR, then the 
dispersion in the Twarog AMR should be at least as large as that of Edv93, not 
smaller. This reinforces the argument that the large scatter in the Edv93 AMR 
is most likely due to selection effects.

Twarog, Ashman, \& Anthony-Twarog (1998)\markcite{taat98} measured a dispersion 
in [Fe/H] of only 0.1 dex for galactic open clusters, which are presumedly less 
subject to diffusion effects than individual field stars. In comparison, the 
average dispersion in the Twarog (1980) AMR is 0.15 dex for field stars with 
ages in the range 3-10 Gyr. If this represents the true dispersion in metallicity
for the field stars, then the difference between the field star and cluster
results (corresponding to a scatter of about 0.1 dex) could be attributable to 
stellar diffusion, although the effect of sampling from different parts of the 
galactic metallicity gradient needs to be accounted for as well. A comparison of 
the cluster data with a complete sample of field stars having kinematic data could
provide a more stringent test of stellar diffusion.

Contamination by thick disk stars complicates the determination of the metallicities
of the oldest thin disk stars. On the other hand, the kinematic data suggest that 
the thick disk may be an ancient population (see Fig. 31 of Edv93 and corresponding
discussion in Freeman 1991 \markcite{kf91}). With larger complete samples of stars 
with metallicity and kinematic measurements it may be possible to subtract the thick 
disk contribution statistically. This would be a very important measurement because 
the initial metallicity of the thin disk is poorly known (as is its age as well).
Determination of this quantity would provide an important constraint on chemical
evolution models.

Abundance measurements for the interstellar medium in galaxies typically imply 
small dispersions in metallicity. Kennicutt \& Garnett (1996) \markcite{kg96} 
found a dispersion of only 0.1-0.2 dex about the radial gradient in O/H from 
41 \hii\ regions in the spiral galaxy M101, consistent with observational 
uncertainties and implying that the intrinisic abundance dispersion is negligible. 
Kobulnicky \& Skillman (1996\markcite{ks96}, 1997\markcite{ks97}) found a dispersion 
in O/H of only $\pm$0.05 dex in the dwarf irregular galaxy NGC 1569, and $\pm$0.10 
dex in NGC 4214. Closer to home, Meyer, Jura, \& Cardelli (1998)\markcite{mjc98} 
have found a very small dispersion, only $\pm$0.05 dex in O/H, in local (within 
500 pc) diffuse interstellar gas. The combined data from ISM and star cluster 
observations imply that the ISM is relatively well-mixed on size scales $<$ 0.5 
kpc and $>$ 1 kpc, or that mixing occurs on sufficiently large spatial and time 
scales that supernova ejecta are considerably diluted by ambient gas. 

Roy \& Kunth (1995)\markcite{rk95} and Elmegreen (1998)\markcite{elm98} discuss 
mixing of SN ejecta on small ($<$ 1 kpc) scales. The implication from these studies 
is that it is difficult to maintain a metallicity dispersion greater than 0.15 dex 
because of the efficiency of mixing processes. Elmegreen (1998), considering the 
enrichment of clouds by supernovae, predicts inhomogeneities of only about 0.05 
dex within molecular clouds. This level of inhomogeneity is indeed consistent with 
the abundance dispersion measured in interstellar gas, but not with the apparent 
dispersion on stellar metallicities. 
%Infall of metal-poor clouds is a popular mechanism to introduce metallicity 
%anomalies. Such events should occur frequently but irregularly to account 
%for both the apparent stellar metallicity dispersion and the small dispersion
%in ISM abundances. 
On the other hand, data on the dispersion in abundances on intermediate size scales 
is lacking. Further studies of interstellar abundances over size scales of $\lesssim$ 
1 kpc, along with further studies of stellar abundances with age, are needed to 
improve our understanding of the distribution and mixing of heavy elements in our 
galaxy and others.

To summarize, we conclude that, while the Edv93 stellar abundance data provide
invaluable information on the evolution of element abundance ratios over time, it 
is not possible to infer either the dispersion in metallicity nor even the shape of 
the age-metallicity relation from these data because of various selection biases, 
as pointed out by the authors themselves. As a result of these biases, we infer
that the intrinsic scatter in the Edv93 AMR must be much smaller than that measured.
We therefore argue that the Twarog (1980) study remains at present the preferred
determination of the solar neighborhood AMR, until a new study based on a complete 
sample of stars becomes available.

\acknowledgments
We thank B. Edvardsson, B. Gustafsson, A. Quillen, and R. Wyse for informative 
discussions, and V. Smith for comments on the manuscript. We also thank the
referee, Bruce Twarog, for a very helpful and stimulating discussion of the 
issues underlying the analysis here. DRG is grateful for support from NASA-LTSARP 
grant NAG5-7734, while HAK acknowledges support from NASA and STScI through 
Hubble Fellowship HF-1094.01-97A.

\clearpage

\begin{figure}
%\plotone{Fig1.ps}
\plotfiddle{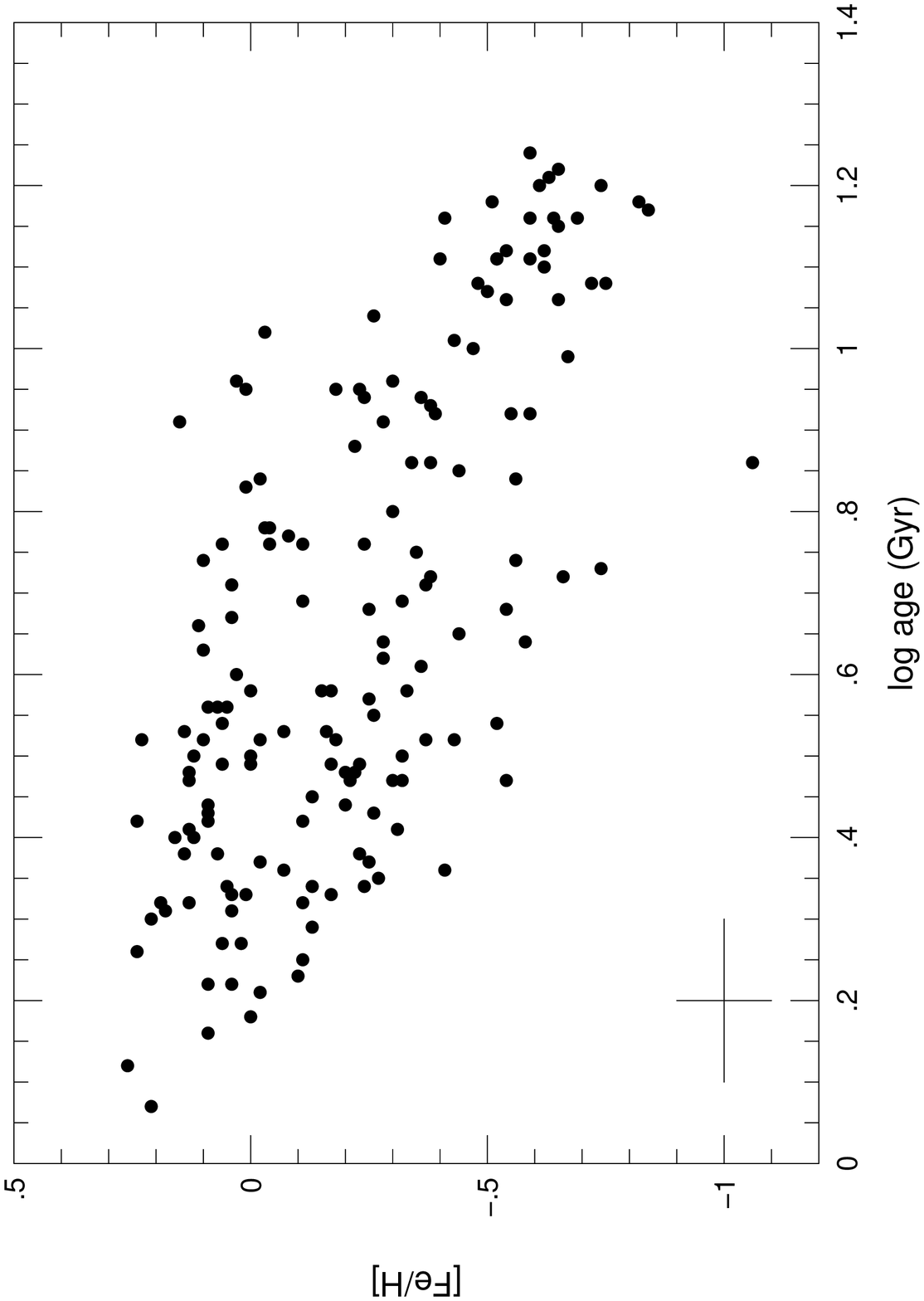}{7in}{270}{70}{70}{-250}{400}
\figcaption[Garnett.fig1.ps]{The revised age vs. metallicity diagram
for stars from the Edvardsson et al. (1993) sample. Stellar ages in 
this case have been taken from Ng \& Bertelli (1998). }
\end{figure}

%\clearpage

\begin{figure}
%\plotone{Fig2.ps}
\plotfiddle{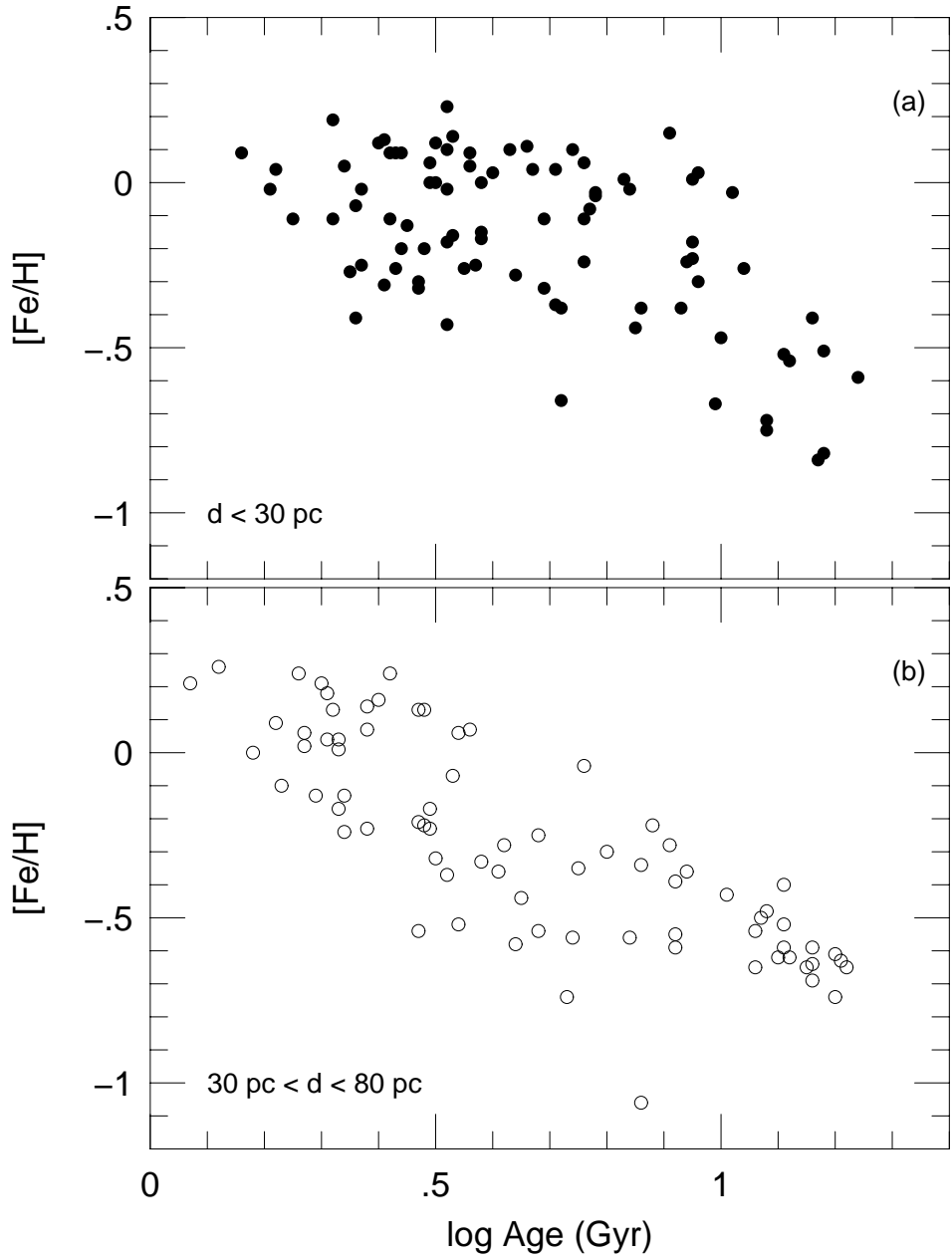}{7in}{00}{70}{70}{-200}{000}
\figcaption[Garnett.fig2.ps]{Age-metallicity relations for Edvardssson 
et al. stars, based on ages from Ng \& Bertelli (1998) and distances
from Hipparcos parallaxes. The sample is divided according to distance. 
{\it (a)} Stars within 30 pc of the Sun. {\it (b)} Stars between 30 and 
80 pc distant. Note the strong asymmetry in the distribution of [Fe/H] 
between the two samples.}
\end{figure}

%\clearpage

\begin{figure}
%\plotone{Fig3.ps}
\plotfiddle{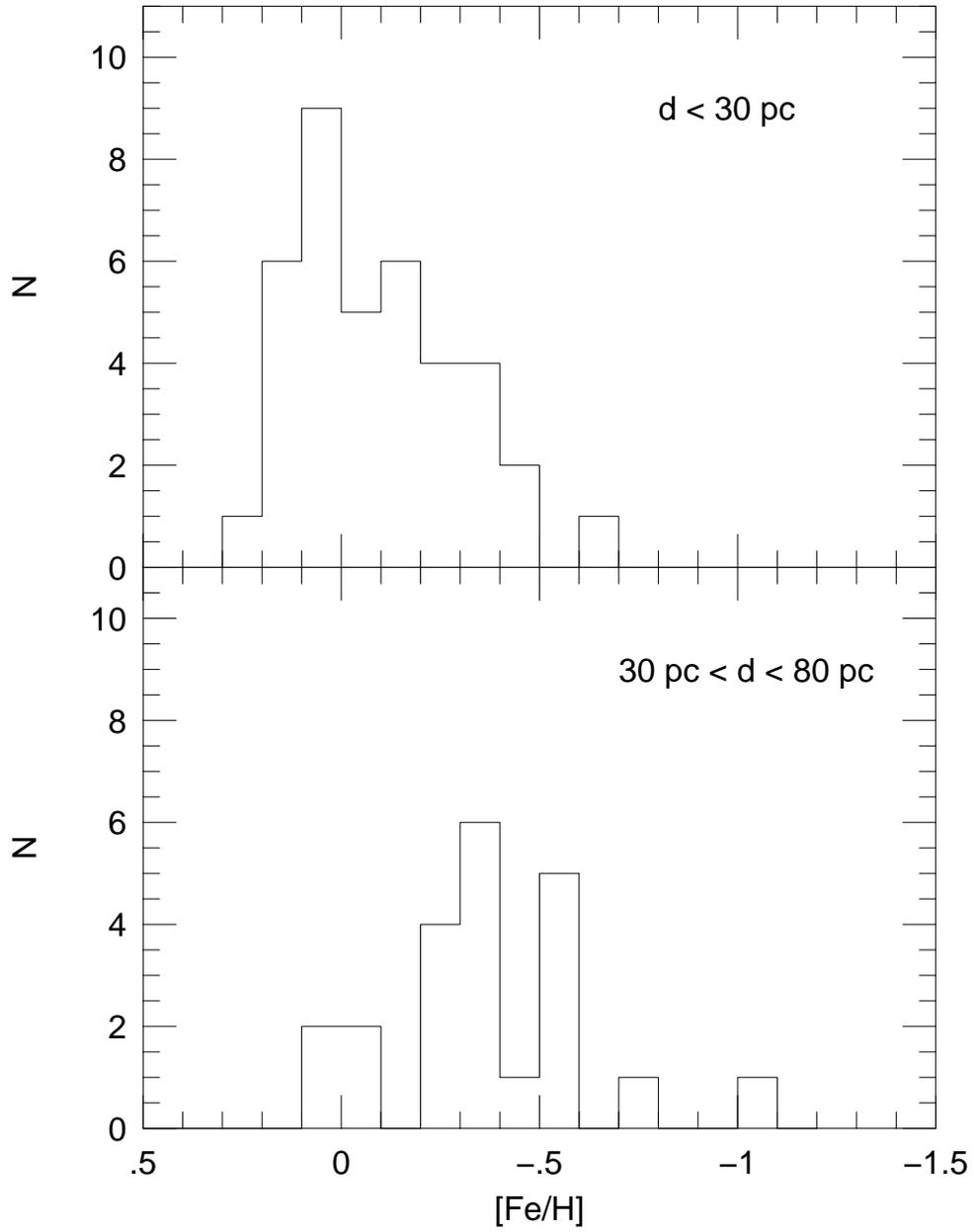}{7in}{00}{70}{70}{-200}{000}
\figcaption[Garnett.fig3.ps]{Histograms of the [Fe/H] distributions 
for stars in the age range 0.5 $<$ log $\tau(Gyr)$ $<$ 0.9. The top
panel shows stars within 30 pc of the sun; the lower panel shows the
stars between 30 and 80 pc.}
\end{figure}

%\clearpage

\begin{figure}
%\plotone{Fig4.ps}
\plotfiddle{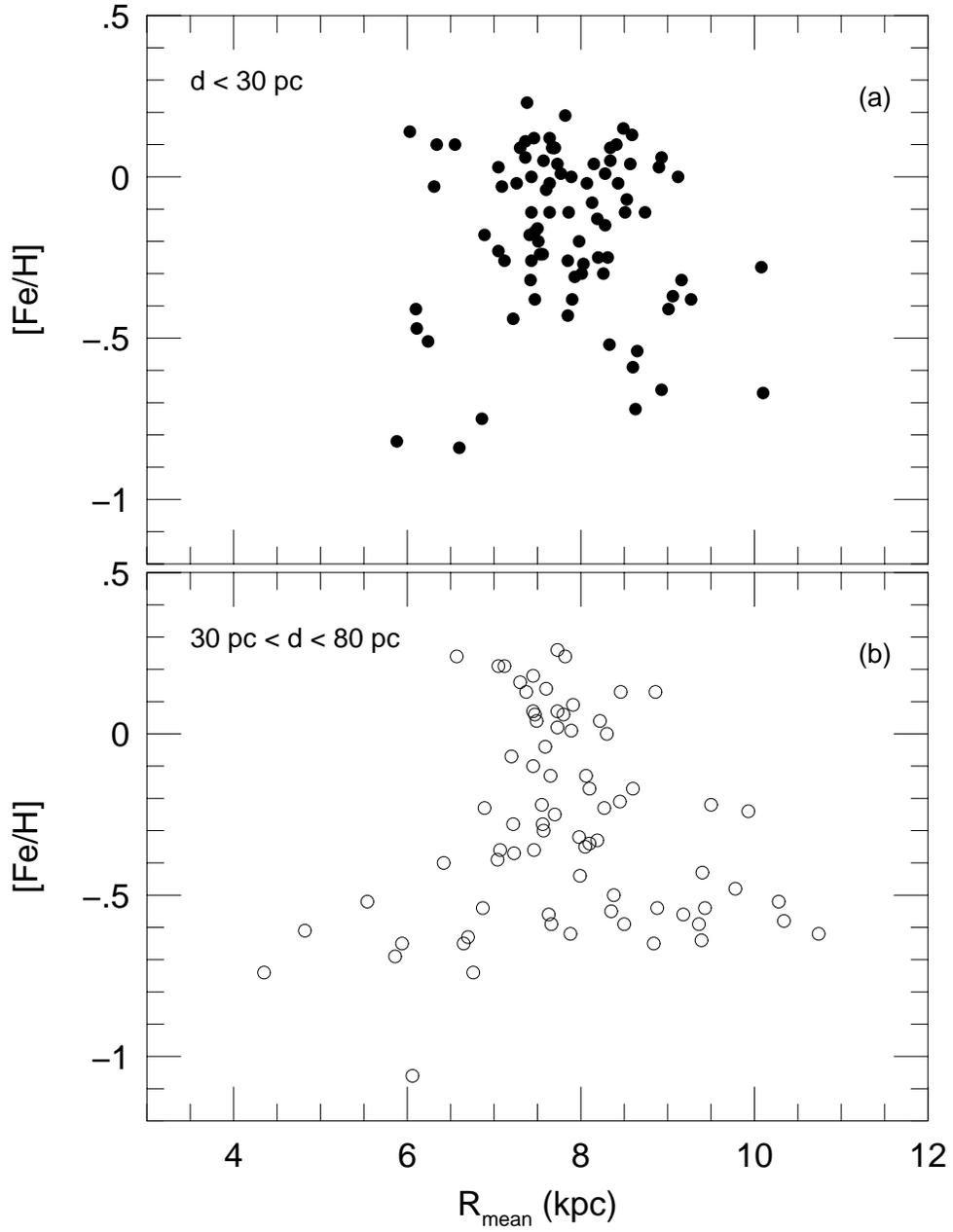}{7in}{00}{70}{70}{-200}{000}
\figcaption[Garnett.fig4.ps]{Stellar [Fe/H] plotted against mean orbital
radius (taken from Edvardsson et al. 1993). Note that many of the metal-poor
stars have orbits that range well beyond the solar circle.}
\end{figure}

%\clearpage

\begin{figure}
%\plotone{Fig5.ps}
\plotfiddle{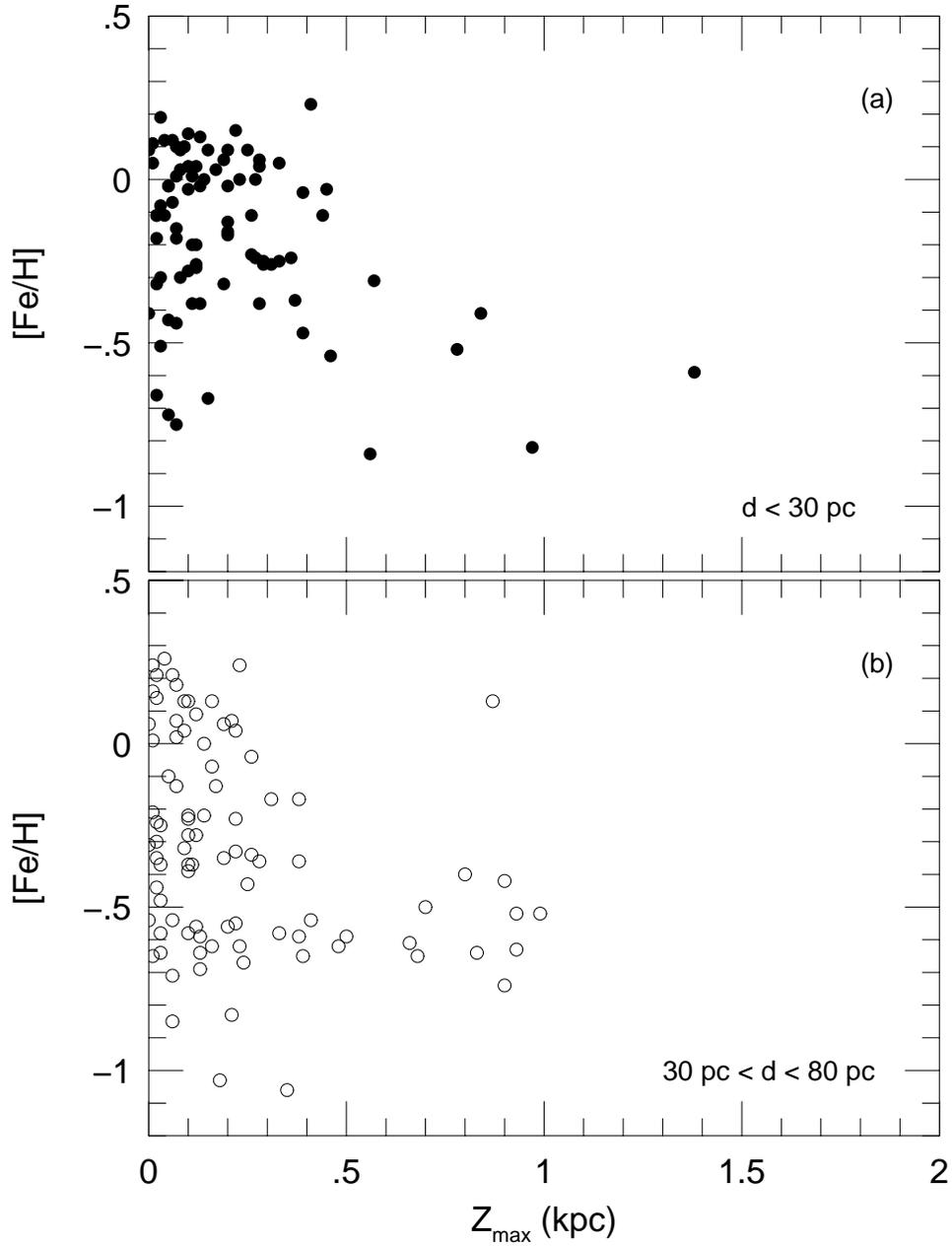}{7in}{00}{70}{70}{-200}{000}
\figcaption[Garnett.fig5.ps]{Stellar [Fe/H] plotted against maximum distance
from the galactic plane (taken from Edvardsson et al. 1993). Many of the
metal-poor stars range far away from the plane. One star, HD148816, lies off
the plot at [Fe/H] = $-$0.74, $Z_{max}$ = 5.44 kpc. }
\end{figure}

%\clearpage

\begin{figure}
%\plotone{Fig6.ps}
\plotfiddle{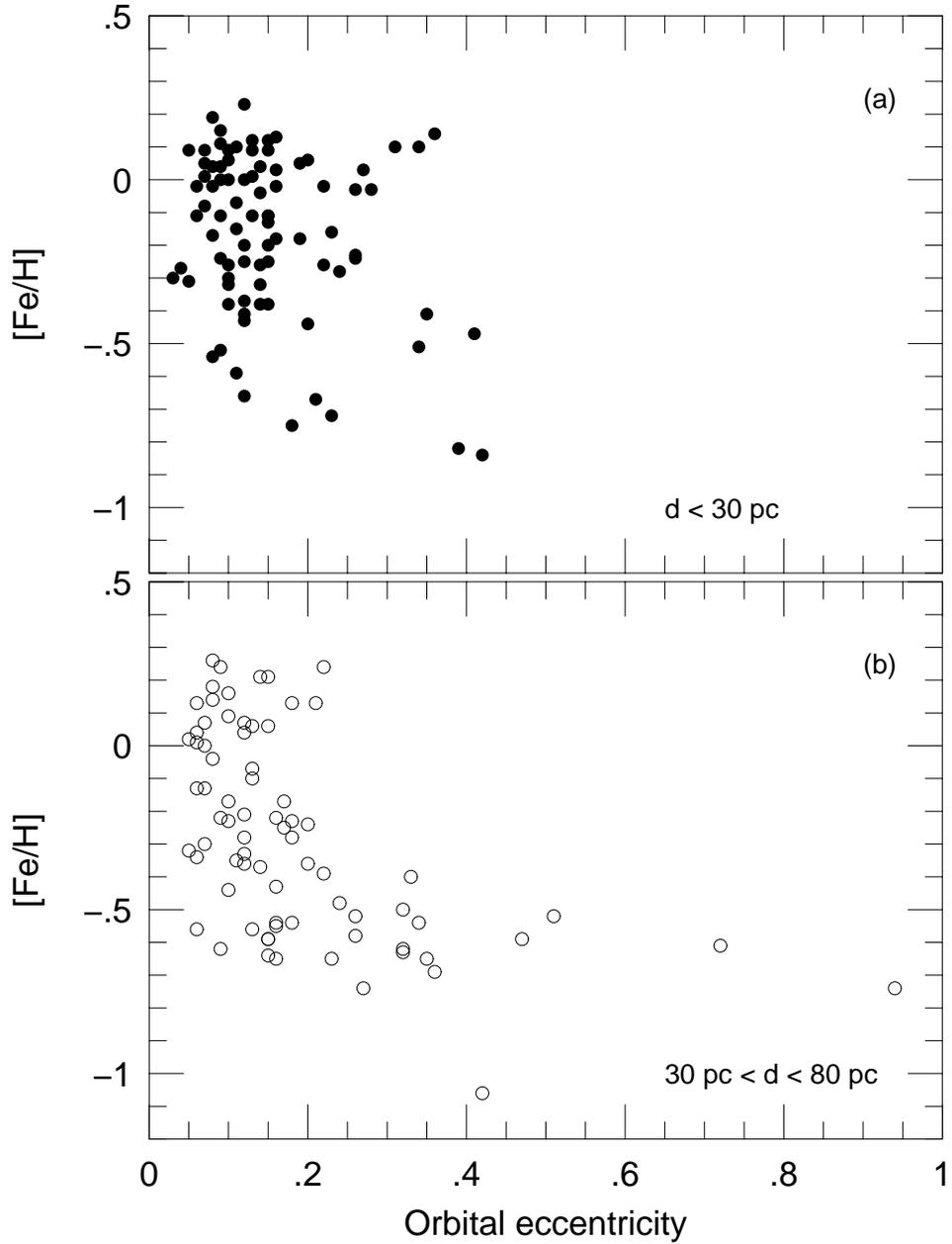}{7in}{00}{70}{70}{-200}{000}
\figcaption[Garnett.fig6.ps]{Stellar [Fe/H] plotted against orbital 
eccentricity $e$ (taken from Edvardsson et al. 1993). Many of the 
metal-poor stars are seen to have highly eccentric orbits.  }
\end{figure}

%\clearpage

\begin{figure}
%\plotone{Fig7.ps}
\plotfiddle{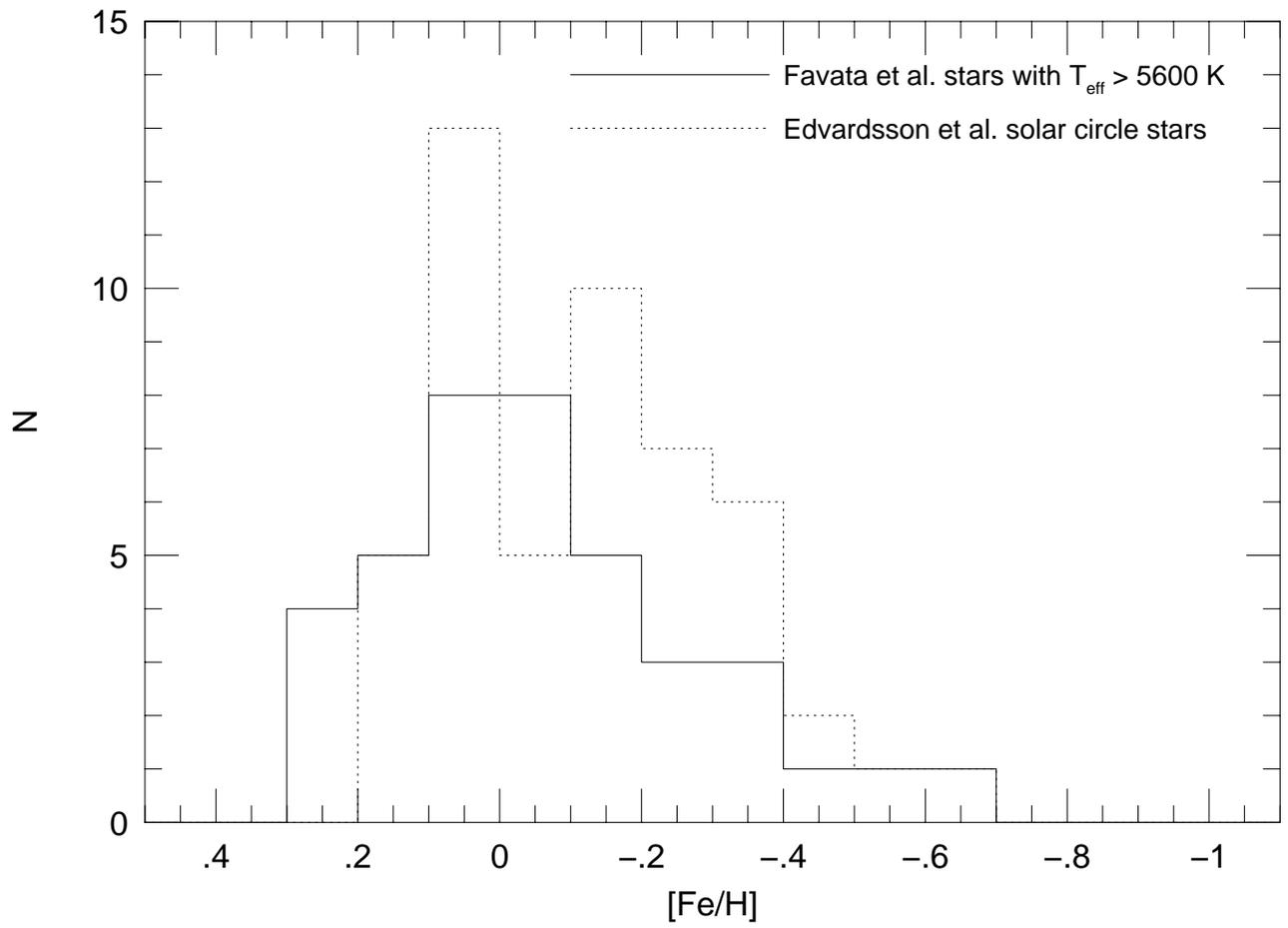}{7in}{270}{70}{70}{-250}{400}
\figcaption[Garnett.fig7.ps]{Comparison of the [Fe/H] distribution
for G dwarfs from Favata et al. (1997) (solid histogram) with that
for the stars in the top panel of Fig. 2 (dotted histogram). }
\end{figure}
%\clearpage

\end{document}